\documentclass[10pt, conference, compsocconf]{IEEEtran}
\usepackage{graphicx}
\usepackage[T1]{fontenc}
\usepackage[]{caption}
\usepackage{amsmath}
\usepackage{array}
\usepackage{fixltx2e}
\usepackage{booktabs}
\usepackage[utf8x]{inputenc}
\usepackage{CJK}

\ifCLASSINFOpdf
\else
\fi
\begin{document}

\title{America Tweets China: A Fine-Grained Analysis of the State and Individual Characteristics Regarding Attitudes towards China}

\author{\IEEEauthorblockN{Yu Wang}
\IEEEauthorblockA{Department of Political Science\\
University of Rochester\\
Rochester, NY, 14627, USA\\
\textit{ywang176@ur.rochester.edu}}
\and
\IEEEauthorblockN{Jianbo Yuan}
\IEEEauthorblockA{Department of Computer Science\\
University of Rochester\\
Rochester, NY, 14627, USA\\
\textit{jyuan10@ur.rochester.edu}}
\and
\IEEEauthorblockN{Jiebo Luo}
\IEEEauthorblockA{Department of Computer Science\\
University of Rochester\\
Rochester, NY, 14627, USA\\
\textit{jluo@cs.rochester.edu}}
}


%


\maketitle

\begin{abstract}
The U.S.-China relationship is arguably the most important bilateral relationship in the 21st century. Typically it is measured through opinion polls, for example, by Gallup and Pew Institute. In this paper, we propose a new method to measure U.S.-China relations using data from Twitter, one of the most popular social networks. Compared with traditional opinion polls, our method has two distinctive advantages. First, our sample size is significantly larger. National opinion polls have at most a few thousand samples. Our data set has 724,146 samples. The large size of our data set enables us to perform state level analysis, which so far even large opinion polls have left unexplored. Second, our method can control for fixed state and date effects. We first demonstrate the existence of inter-state and inter-day variances and then control for these variances in our regression analysis. Empirically, our study is able to replicate the stylized results from opinion polls as well as generate new insights. At the state level, we find New York, Michigan, Indiana and Arizona are the top four most China-friendly states. Wyoming, South Dakota, Kansas and Nevada are most homogeneous. At the individual level, we find attitudes towards China improve as an individual's Twitter experience grows longer and more intense. We also find individuals of Chinese ethnicity are statistically more China-friendly. 

\end{abstract}


%
\IEEEpeerreviewmaketitle

\section{Introduction}
In international relations, perceptions matter \cite{Perceptions,Trust}. And how the U.S. perceives China matters particularly, the former being the world's only superpower and the latter a potential challenger. Typically these perceptions are measured using opinion polls. For example, a February 2015 survey by \textit{Foreign Policy} shows that the majority of American students who studied in China have developed a more positive view of the country (sample size: 343) \cite{StudyChina}. A February 2014 poll survey by \textit{Gallup} shows that 53\% of Americans view China very or mostly unfavorably (sample size: 1,023) \cite{ViewChinaUnfavorably}.




While opinion polls have been the standard way for gauging public opinion, their weaknesses are obvious. First and foremost, the small sample size, as evidenced above, renders any fine grained analysis extremely difficult if not impossible \cite{stateLevel}. Second, these opinion polls, mostly carried out on an annual basis, are susceptible to the influence of daily events. Surveys before or after the event can yield drastically different results, and yet both will be used as representing the annual result. In this paper, we propose a method that can solve these two problems. We measure U.S. perceptions of China through Twitter.

With 302 million active monthly users, Twitter is one of the most popular social networks in the world.\footnote{https://about.twitter.com/company.} In the U.S. alone there are 69.46 million users. The huge amount of data generated by U.S. users is an ideal repository for mining U.S. perceptions of China. It enables us to achieve what have so far evaded opinion polls. Specifically, with a large data set, we are able to carry out state level analysis as never before. By utilizing the time-stamps contained in tweets, we are able to control for fixed time effects. Utilizing the location information, we are able to control for fixed state effects.



\begin{figure}[h!]
\includegraphics[height=6cm, width=9cm]{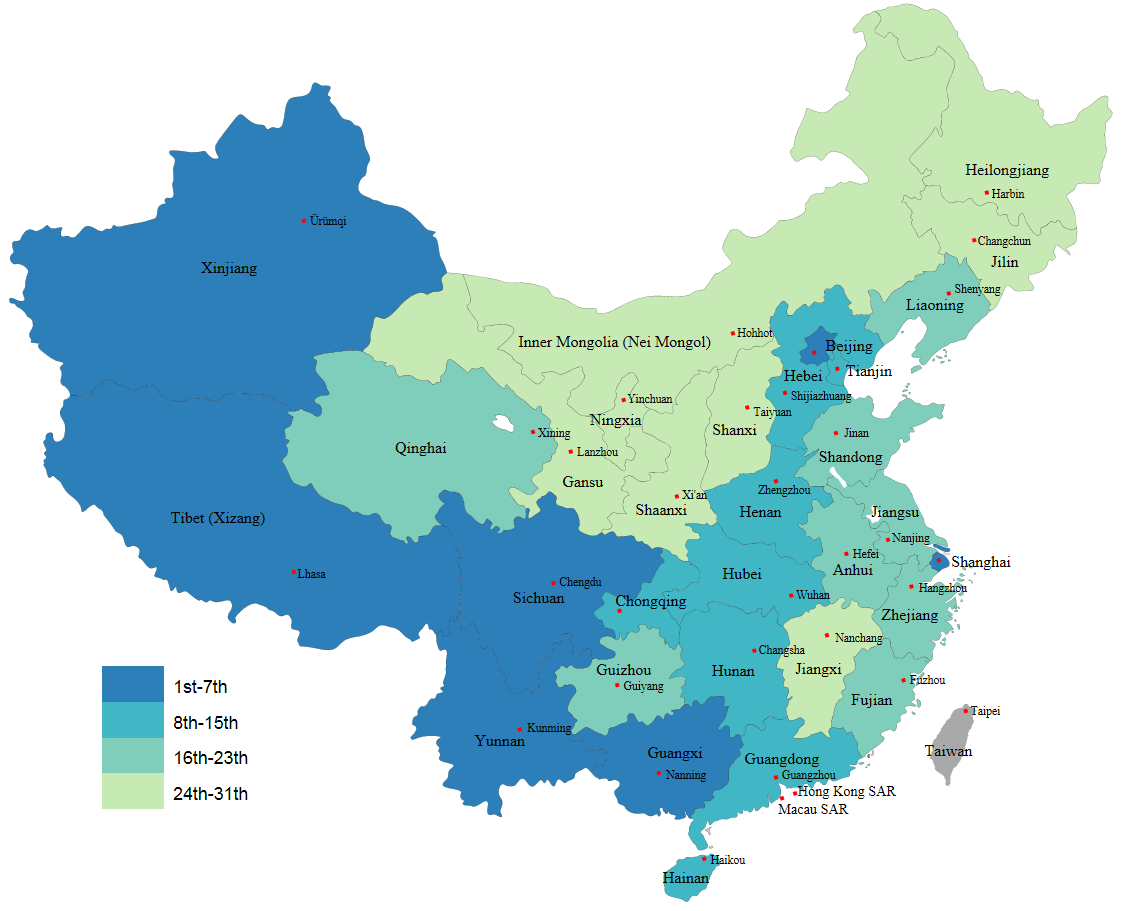}
\caption{America Tweets China, aggregated at the provincial level. This map is generated with 25,677 China-focused tweets. Data does not include Hong Kong SAR, Macao SAR or Taiwan.}
\end{figure}

Empirically, our study successfully replicates the stylized findings from conventional opinion polls as well as generate new insights. We find that New York, Michigan, Indiana and Arizona are the top four China-friendly U.S. states and that Wyoming, Wisconsin, South Dakota and West Virginia are the least friendly. Wyoming, South Dakota, Kansas and Nevada are the most homogeneous in attitudes towards China. Michigan, New Hampshire, New Jersey and Wisconsin are the least homogeneous. At the individual level, we find attitudes towards China improve as the individuals' Twitter experience grows longer and more intense. We also find that individuals of Chinese ethnicity are statistically more China-friendly.

Our paper proceeds as follows. Section 2 presents related literature on international relations, sentiment analysis using Twitter data, and inferring geo-information in the tweets. Section 3 presents our data and data processing procedures. Section 4 presents our state-level analysis. Section 5 presents the individual level analysis. Section 6 concludes.

\section{Literature Review}

Our study builds on previous research both in international relations and in computer science. 

China's rise is quickly reshaping the post-Cold War international structure \cite{Structuralism}. Lake argues that if China continues to grow, it is likely to bid for its own subordinates to counter America's current hierarchies \cite{hierarchy}. Mearsheimer, a proponent of offensive realism, contends that China's rise will not be peaceful and that China will ``try to dominate Asia the way the United States dominates the Western Hemisphere" \cite{Mearsheimer2014}. 

Perceptions matter in international relations \cite{Perceptions,Trust} and particularly so in the relations between the U.S. and China. Johnson, through analyzing Chinese publications, argues that the common description in U.S. media, pundit, and academia of an increasingly assertive China is ill-founded and points out the dangers of misperception \cite{Assertiveness}. Accurate perceptions are likely to contribute to trust while misguided perceptions could well lead to conflict and even war.


We believe tweets can be used to measure U.S. perceptions. The abundance of data generated through Twitter has attracted researchers from various fields. Bollen et al. use Twitter mood to predict the stock market \cite{stock_price}; Paul and Drezde mine tweets for public health topics \cite{publicHealth}; and An et al. use Twitter data to track opinions about climate change \cite{climateChange}. Among this group of researchers there are also political scientists. For example, Tumasjan et al. use the tweets to predict election results \cite{tumasjan2010predicting}. Barberá analyzes the network structure of Twitter users to infer political ideology \cite{Ideology}.

In order to perform state level analysis, our study makes extensive use of the geo-information in the tweets. In this effort, we benefit from the research by Hecht et al. \cite{geolocation}. Their work identifies various problems with geo-information. We adhere to their suggestions and select only those tweets that have a case-sensitive state address such as \textit{CA} in ``Los Angeles, CA", \textit{New York} in "Upper Manhattan, New York" and \textit{Vermont} in ``Vermont, USA." For the purpose of this study, we stop at the state level and do not go to the city level. 

\section{Data and Processing}

In this section, we first describe our data set and the processing procedures. Second, we define tweet level, state level and user level features respectively. Third, we use time-series tweets and pseudo-labeled tweets to test the validity of the sentiment analysis tool TextBlob.

\subsection{Data}
We compile a corpus of tweets using the Twitter search API between 10th and 29th of March, and between May 16th and June 15th. We perform a query-based search (\textit{China OR Chinese}) to collect English tweets related to China. We then select those tweets with a state address. When a tweet can be attributed to multiple U.S. states, we attribute that tweet to all the identified states. Details of the processing procedure are described below. In the end, after removing duplicates, we have collected 724,146 tweets, each associated with a unique tweet id, a state id, and a time stamp.\footnote{The data sets and the codes are available at the authors' website: https://sites.google.com/site/wangyurochester/.}
\rule{8.6cm}{0.4pt}


\noindent\textbf{Input}: tweet\textsubscript{i}

\noindent\textbf{Output}: tweet\textsubscript{i}'s attributes, or null

\noindent\textbf{If} tweet\textsubscript{i} does not contain stop words

\textbf{For} each state\textsubscript{j} $\in$  \{51 U.S. states\}\footnote{For the purpose of this study, we treat Washington, D.C. as a state.}.

\hspace{0.5cm} \textbf{If} state\textsubscript{j} in tweet\textsubscript{i}.\textit{place}

\hspace{1cm} assign tweet\textsubscript{i} to state\textsubscript{j}

\hspace{1cm} extract all the attributes

\hspace{1cm} evaluate polarity of tweet\textsubscript{i}

\hspace{1cm} \textbf{For} each province\textsubscript{k} $\in$  \{31 Chinese provinces\}

\hspace{1.5cm} \textbf{If} province\textsubscript{k} in tweet\textsubscript{i}.\textit{text}

\hspace{2cm} assign tweet\textsubscript{i} to province\textsubscript{k}

\hspace{1cm} \textbf{For} each name\textsubscript{l} $\in$ \{100 common Chinese names\}

\hspace{1.5cm} \textbf{If} name\textsubscript{l} in tweet\textsubscript{i}.\textit{username}

\hspace{2cm} Chinese=1

\hspace{1.5cm} \textbf{Else}

\hspace{2cm} Chinese=0

\hspace{0.5cm} \textbf{End if}

\textbf{End for}

\noindent\textbf{End if} 

\noindent\rule{8.6cm}{0.4pt}


From China's Xinhua News Agency's website, we obtain the top 100 most common Chinese surnames, which are used by 84.77\% of the Chinese population.\footnote{http://news.xinhuanet.com/society/2007-04/24/content\_6021482.htm.}  We use these names to identify Twitter users of Chinese ethnicity. When translated into \textit{pinyin}, these 100 names result in 85 distinct names. We report these names in alphabetic order in Table 1. 

\begin{table}[h]
\centering
\caption{Most Common Chinese Surnames}
\label{my-label}
\begin{tabular}{lllll}
Bai   & Cai                     & Cao  & Ceng  & Chen \\
Cheng & Cui                     & Dai  & Deng  & Ding \\
Dong  & Du                      & Duan & Fan   & Fang \\
Feng  & Fu                      & Gao  & Gong  & Gu   \\
Guo   & Han                     & Hao  & He    & Hou  \\
Hu    & Huang                   & Jia  & Jiang & Jin  \\
Kong  & Lei                     & Li   & Liang & Liao \\
Lin   & Liu                     & Long & Lu    & Luo  \\
Lv    & Ma                      & Mao  & Meng  & Mo   \\
Pan   & Peng                    & Qian & Qin   & Qiu  \\
Ren   & Shao                    & Shi  & Song  & Su   \\
Sun   & Tan                     & Tang & Tao   & Tian \\
Wan   & Wang                    & Wei  & Wu    & Xia  \\
Xiang & Xiao                    & Xie  & Xiong & Xu   \\
Xue   & Yan                     & Yang & Yao   & Ye   \\
Yin   & Yu                      & Yuan & Zhang & Zhao \\
Zheng & Zhong                   & Zhou & Zhu   & Zou 
\end{tabular}
\end{table}

Out of the 101,907 individuals in the sample data set, we are able to identify 938 individuals with Chinese surnames, which represents 0.923\% of the sample. This is close to the official figure 1.02\%, published by the Census Bureau in 2010.\footnote{http://www.census.gov/prod/cen2010/briefs/c2010br-11.pdf.} 
\subsection{Features}
\noindent Tweet features:

\noindent \textbf{Polarity:} This is defined as how positive the tweet is, ranging from -1 to 1. Polarity is calculated using Textblob. With polarity, we can view each China-focused tweet as a vote.


\:

\noindent Other tweet features include followers, followees, retweets, and reply (binary). 

\:

\noindent State level features:

\noindent \textbf{Friendliness:} This is defined as the arithmetic average of the tweets' polarity scores for each state, ranging from -1 to 1. $F_{s}=\frac{\sum f_{s,i}}{n}$. It measures the state's aggregate sentiment towards China.

\noindent \textbf{Variance:} This is defined as the variance of the tweets' polarity scores for each state. $V_{s}=Var(f_{s,i})$. Variance measures how varied each state's attitudes are towards China. We call a state homogeneous if the variance is small. 

\:

\noindent Individual level features:

\noindent \textbf{Experience:} This is defined as the length of the period the individual has been using Twitter. It is calculated as date\textsubscript{1}-date\textsubscript{0}, where date\textsubscript{1} is the day when the tweet is posted and date\textsubscript{0} is the day when the Twitter account is created. 

\noindent \textbf{Intensity:} This is defined as the average number of tweets the individual posts per day. It is calculated as $\frac{\# tweets}{date\textsubscript{1}-date\textsubscript{0}+1}.$

\noindent \textbf{Chinese:} This is defined as whether the individual has a Chinese surname. It is binary. 

\subsection{External Validity}
 
We validate the viability of using Textblob to measure sentiments towards China.\footnote{http://textblob.readthedocs.org/en/dev.} So far as we know, there have been no state-level (cross-sectional) opinion polls on U.S.-China relations. All the data measuring U.S. attitudes towards China are limited to the national level. Indeed, this is one of the motivations for this study. We decide to use time series data for the purpose of validation, as significant events between the two countries are easily recognizable and unanimity is easy to achieve.

Three events stood out between May and June: the South China Sea crisis starting on May 20th, the Yangtze River accident on June 1st and Hong Kong protests on June 14th. As shown in Fig. 2 below, these events are well reflected in the aggregate national sentiments. Thus, Textblob passes our time-series test.

\begin{figure}[h!]
\includegraphics[height=6cm, width=8.5cm]{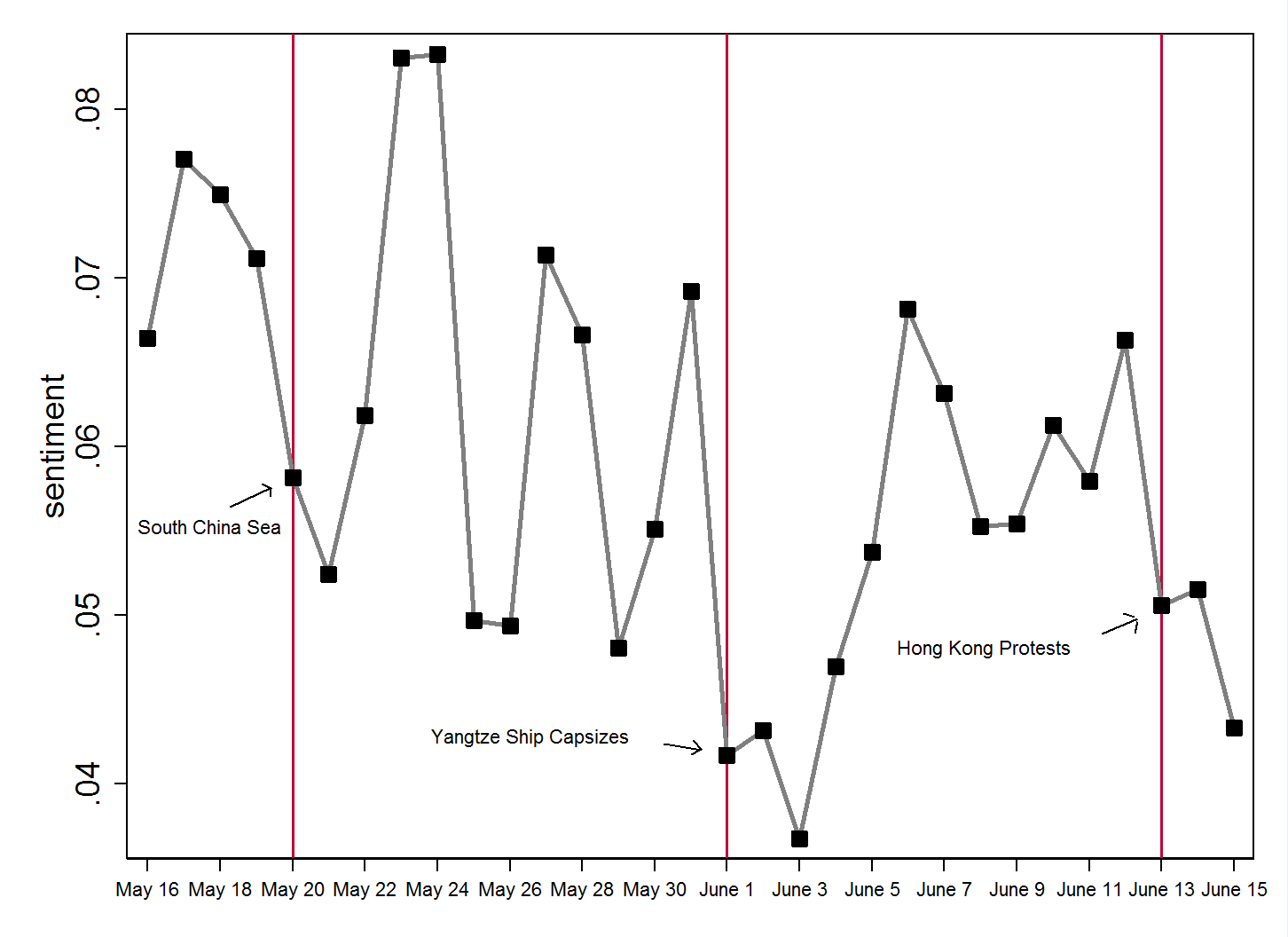}
\caption{U.S. sentiment towards China, May 14-June 15.}
\end{figure}   

We further test Textblob's performance with tweets that contain emoticons. Davidov et al. have shown that smileys, as well as hashtags, in tweets can be used as labels \cite{smiley}. We first choose two specific smileys: ``:)" for positve and ``:(" for negative. We are able to find 1255 tweets with these emoticons in our sample. The testing rule is as follows:

Correct:

\hspace{1cm} if Textblob returns polarity>=0 for ``:)"

\hspace{0.6cm} \textit{or} if Textblob returns polarity<=0 for ``:("

This is a lenient test as tweets that contain emoticons and are marked as neutral are automatically classified as correct. The testing results, reported below, are satisfactory.

{
\hfill

\centering
\begin{tabular}{rl}
\hline
Sample Size    & 1255 \\
Accuracy (\%)  & 92.4 \\
Sensitivity (\%)  &95.6 \\
Specificity (\%) &72.3 \\
\hline
\end{tabular}

\hfill
}

\section{State level Analysis}
In this section, we investigate the state characteristics of the tweets. We first calculate the volume of tweets that can be attributed to each state and compare our statistics with Google Trend. We then create a \textbf{State-Province Matrix} that projects tweets generated in a U.S. state to a Chinese province. The dimension of our matrix is 51 (states) x 31 (provinces). Third, we evaluate the friendliness and variance of each state based on our data set.

\subsection{Volume of Tweets by State}
Following the processing procedure described in Section 3, we obtain 724,146 China-focused tweets geocoded to the state level. We then calculate the total number of tweets assigned to each state. The summary statistics are reported below. 


\begin{table}[!htbp]\centering \caption{Summary statistics per state \label{sumstat}}
\begin{tabular}{l c c c c c}\hline
\multicolumn{1}{c}{\textbf{Variable}} & \textbf{Mean}
 & \textbf{Std. Dev.}& \textbf{Min.} &  \textbf{Max.} & \textbf{N}\\ \hline
tweets & 14199 & 23085 & 1836 & 149043 & 51\\
\hline
\end{tabular}
\end{table}


In terms of the total number of tweets, the top four states are New York, California, Washington, D.C., and Texas. When controlling for state population, Washington, D.C. generates by far the most China-focused tweets per capita. The bottom four states are Vermont, New Mexico, West Virginia and South Dakota. Detailed geographical comparisons are reported in Fig. 3.

\vspace*{0.5\baselineskip}

\begin{figure}[!htbp]
\includegraphics[height=5.5cm, width=8.5cm]{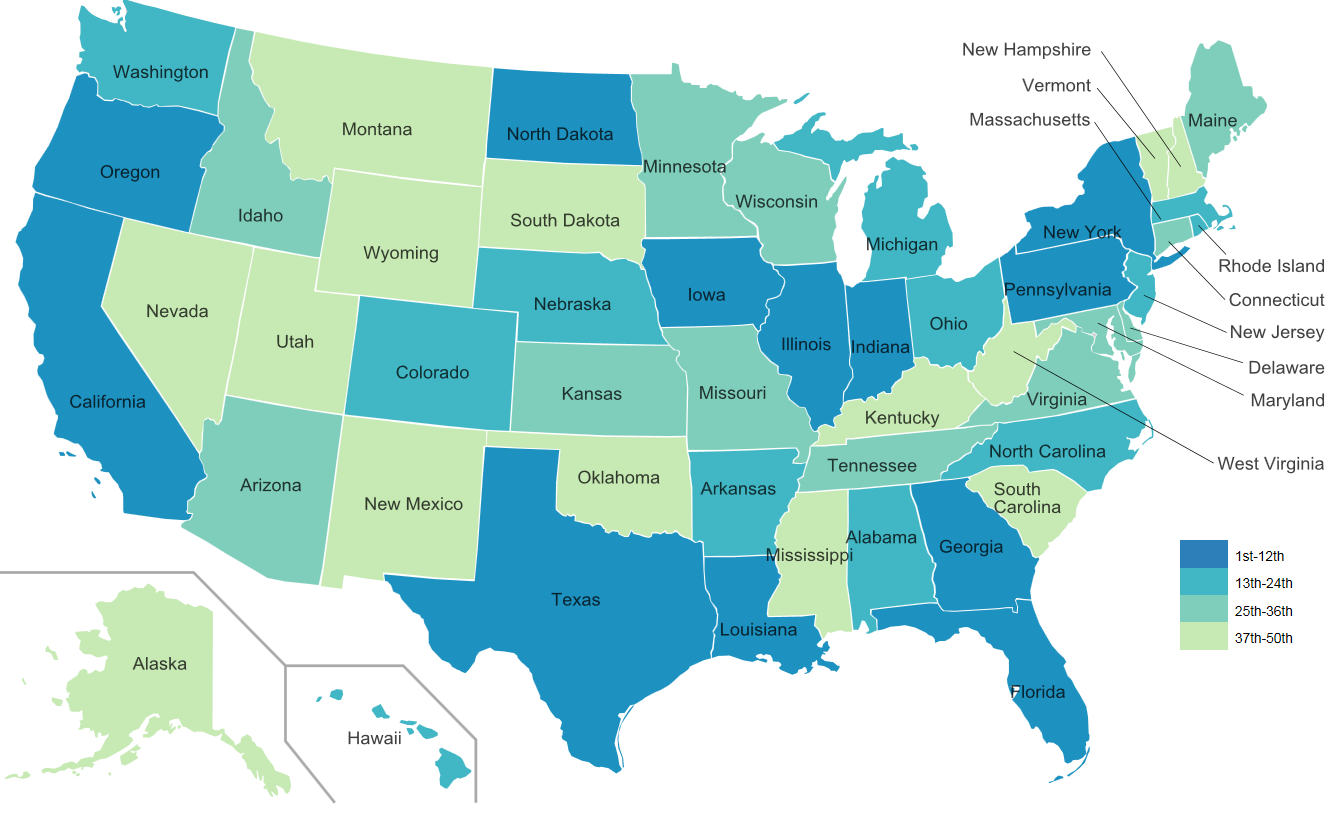}
\caption{America tweets China, aggregated at state level. This map is generated with 724,146 China-focused tweets. Washington, D.C. is not shown on the map.}
\end{figure}
For cross validation, we compare our results with the state-based index generated from the Google Trend.\footnote{https://www.google.com/trends/} The Google index measures the frequency with which people in each U.S. state search for the keyword \textit{China} between Jan. 4, 2004 and Jun. 21, 2015. To make for easy comparison, we first log-transform our counts of tweets. The Google index is used as it is. We plot the Google index as the x axis and plot our Twitter index as the y axis. The result is reported in Fig. 4.

The Twitter index and the Google index are highly correlated, with a correlation coefficient of 0.76 (sample size: 51). One pattern stands out here. The top three states New York, California and Washington, DC score very high by both measures. The bottom four states South Dakota, Montana, Wyoming and New Mexico score very low by both measures. The remaining forty-four states lie in between.

\begin{figure}[h!]
\includegraphics[height=6.5cm, width=8.5cm]{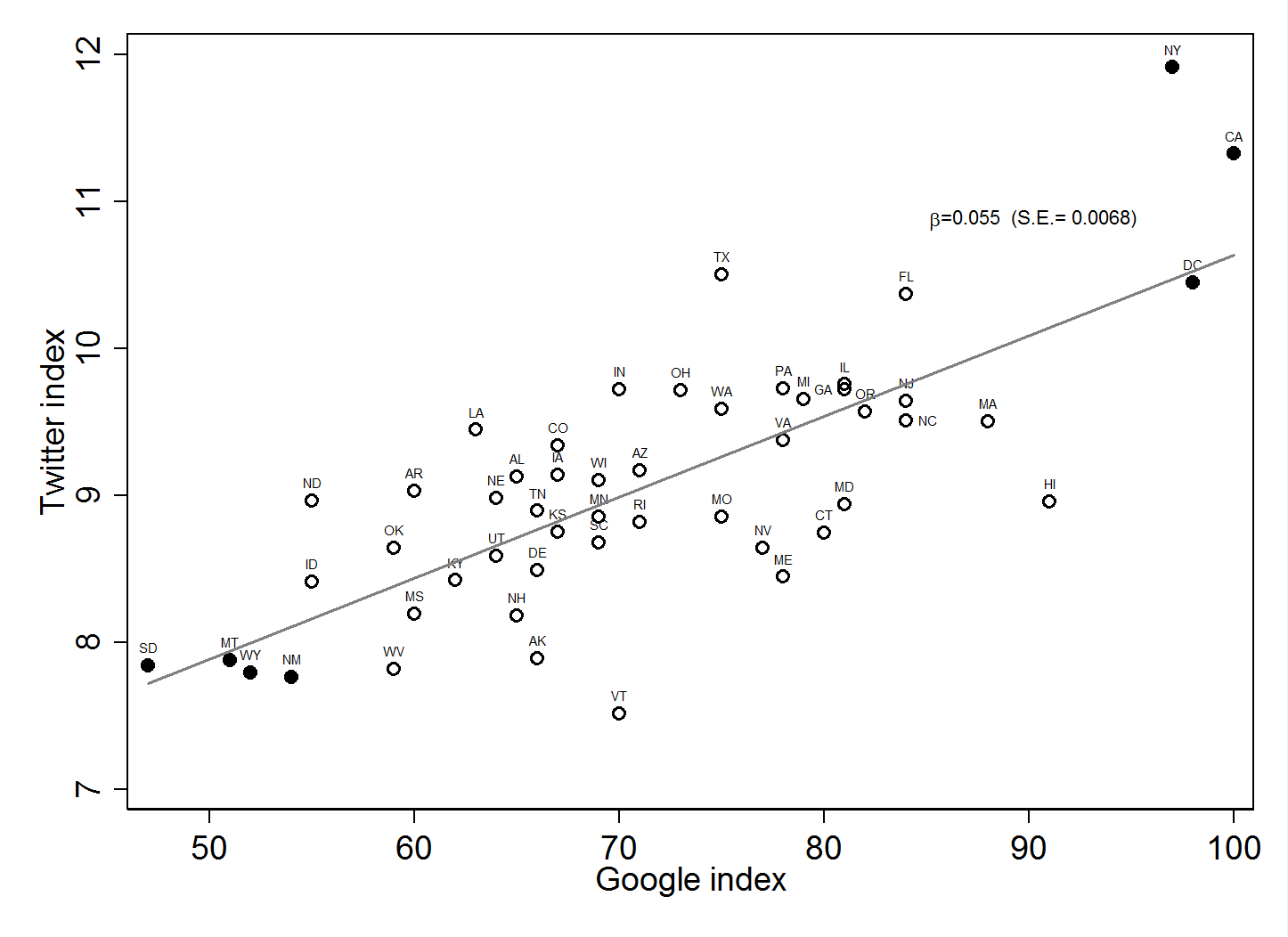}
\caption{Compare Twitter with Google. $\beta$ represents the linear regression coefficient. Standard error of the estimation is reported in parentheses.}
\end{figure}



\subsection{The State-Province Matrix}
The large size of our data set enables us to achieve state level analysis that has so far evaded most researchers. Moreover, by examining the contents of these tweets more closely, we are able to identify which Chinese province a tweet is targeted at. Connecting the state of origin to the province of destination, we can build a State-Province matrix of dimension 51 $\times$ 31. The matrix, built with 25,677 tweets, is reported in Table 3. Each row represents a U.S. state and each column a Chinese province. Values in each row represent the distribution of tweets in the 31 Chinese provinces. Their sum has been normalized to 1. 

\begin{table*}[!htbp]
  \centering
  \tabcolsep=0.08cm
  \renewcommand{\arraystretch}{1.1}

  \caption{State-Province Matrix}
    \begin{tabular}{crrrrrrrrrrrrrrrrrrrrrrrrrrrrrrr}
    \toprule
   Matrix & \begin{CJK}{UTF8}{gkai}皖\end{CJK}&\begin{CJK}{UTF8}{gkai}京\end{CJK} & \begin{CJK*}{UTF8}{gkai}渝\end{CJK*} & \begin{CJK*}{UTF8}{gkai}闽\end{CJK*} & \begin{CJK}{UTF8}{gkai}甘\end{CJK} & \begin{CJK}{UTF8}{gkai} 粤\end{CJK} & \begin{CJK}{UTF8}{gkai}桂\end{CJK} & \begin{CJK}{UTF8}{gkai} 贵\end{CJK} & \begin{CJK}{UTF8}{gkai} 琼\end{CJK} & \begin{CJK}{UTF8}{gkai} 冀\end{CJK} & \begin{CJK}{UTF8}{gkai} 黑\end{CJK} & \begin{CJK}{UTF8}{gkai} 豫\end{CJK} & \begin{CJK}{UTF8}{gkai} 鄂\end{CJK} & \begin{CJK}{UTF8}{gkai} 湘\end{CJK} & \begin{CJK}{UTF8}{gkai} 蒙\end{CJK} & \begin{CJK}{UTF8}{gkai} 苏\end{CJK} & \begin{CJK}{UTF8}{gkai} 赣\end{CJK} & \begin{CJK}{UTF8}{gkai} 吉\end{CJK} & \begin{CJK*}{UTF8}{gkai}辽\end{CJK*} & \begin{CJK}{UTF8}{gkai} 宁\end{CJK} & \begin{CJK}{UTF8}{gkai} 青\end{CJK} & \begin{CJK}{UTF8}{gkai} 陕\end{CJK} & \begin{CJK}{UTF8}{gkai} 鲁\end{CJK} & \begin{CJK}{UTF8}{gkai} 沪\end{CJK} & \begin{CJK}{UTF8}{gkai} 晋\end{CJK} & \begin{CJK}{UTF8}{gkai} 川\end{CJK} & \begin{CJK}{UTF8}{gkai} 津\end{CJK} & \begin{CJK}{UTF8}{gkai} 藏\end{CJK} & \begin{CJK}{UTF8}{gkai} 新\end{CJK} & \begin{CJK}{UTF8}{gkai} 云\end{CJK} & \begin{CJK}{UTF8}{gkai} 浙\end{CJK} \\
    \midrule
    AK & 1.5 & \textbf{29.2} & 1.5 & 0.0 & 0.0 & 0.0 & 0.0 & 0.0 & 0.0 & 1.5 & 0.0 & 3.1 & 0.0 & 0.0 & 0.0 & 0.0 & 0.0 & 0.0 & 0.0 & 0.0 & 0.0 & 0.0 & 0.0 & \textbf{23.1} & 0.0 & 4.6 & 0.0 & \textbf{29.2} & 4.6 & 1.5 & 0.0 \\
    AL & 0.3 & \textbf{39.5} & 1.3 & 0.0 & 0.0 & 1.0 & 3.5 & 0.3 & 0.8 & 1.0 & 0.0 & 1.8 & 1.8 & 1.3 & 0.0 & 0.5 & 0.3 & 0.0 & 0.0 & 0.0 & 0.3 & 0.0 & 0.3 & \textbf{18.9} & 0.0 & 2.5 & 3.8 & \textbf{16.6} & 2.5 & 1.8 & 0.3 \\
    AR & 0.3 & \textbf{41.1} & 1.7 & 0.0 & 0.9 & 2.0 & 2.3 & 0.0 & 0.3 & 0.6 & 0.0 & 0.6 & 1.4 & 1.1 & 0.6 & 0.3 & 0.0 & 0.3 & 0.6 & 0.0 & 0.3 & 0.9 & 0.0 & \textbf{22.6} & 0.0 & 1.1 & 2.3 & \textbf{14.0} & 3.4 & 1.4 & 0.0 \\
    AZ & 0.9 & \textbf{50.0} & 0.9 & 0.9 & 0.0 & 2.7 & 2.7 & 0.6 & 1.8 & 0.0 & 0.0 & 1.2 & 0.9 & 0.6 & 0.3 & 0.0 & 0.3 & 0.0 & 0.6 & 0.0 & 1.2 & 0.0 & 0.3 & \textbf{19.3} & 0.0 & 1.5 & 1.8 & \textbf{8.4} & 1.8 & 1.2 & 0.0 \\
    CA & 0.3 & \textbf{37.0} & 1.0 & 0.2 & 0.2 & 0.8 & \textbf{14.6} & 0.4 & 1.6 & 0.6 & 0.1 & 0.9 & 1.0 & 2.0 & 0.0 & 0.5 & 0.1 & 0.1 & 0.4 & 0.2 & 0.2 & 0.2 & 0.5 & \textbf{19.9} & 0.2 & 1.7 & 0.9 & \textbf{10.3} & 2.7 & 1.3 & 0.3 \\
    CO & 0.4 & \textbf{35.3} & 1.7 & 0.0 & 0.2 & 0.4 & 2.1 & 0.4 & 0.6 & 0.6 & 0.0 & 1.3 & 1.3 & 1.9 & 0.0 & 0.4 & 0.2 & 0.2 & 0.6 & 0.0 & 0.4 & 0.0 & 0.2 & \textbf{19.3} & 0.0 & 1.5 & 0.6 & \textbf{14.1} & \textbf{13.1} & 2.8 & 0.0 \\
    CT & 1.1 & \textbf{36.7} & 0.7 & 0.4 & 0.0 & 2.6 & 0.7 & 0.0 & 1.5 & 4.9 & 0.0 & 0.0 & 0.4 & 0.0 & 0.0 & 0.0 & 0.0 & 0.0 & 0.0 & 0.0 & 0.0 & 0.0 & 1.9 & \textbf{25.1} & 0.0 & 1.5 & 4.1 & \textbf{13.9} & 3.7 & 0.4 & 0.4 \\
    DC & 0.1 & \textbf{50.9} & 1.0 & 0.5 & 0.1 & 0.8 & 0.4 & 0.1 & 1.5 & 0.5 & 0.1 & 0.5 & 0.7 & 0.5 & 0.1 & 1.0 & 0.3 & 0.2 & 1.0 & 0.1 & 0.5 & 0.4 & 0.5 & \textbf{15.2} & 0.2 & 0.7 & 1.0 & \textbf{12.0} & \textbf{7.6} & 1.2 & 0.5 \\
    DE & 0.7 & \textbf{24.3} & 0.7 & 0.7 & 1.4 & 0.7 & 0.7 & 0.7 & 0.7 & 1.4 & 0.0 & 1.4 & 1.4 & 0.7 & 0.0 & 0.7 & 0.0 & 0.0 & 0.7 & 0.7 & 0.7 & 0.7 & 0.7 & \textbf{26.4} & 0.0 & 2.0 & 0.0 & \textbf{21.6} & \textbf{9.5} & 0.7 & 0.7 \\
    FL & 0.1 & \textbf{43.3} & 1.1 & 0.3 & 0.1 & 0.3 & 2.5 & 0.0 & 1.8 & 0.3 & 0.2 & 1.3 & 1.0 & 1.4 & 0.1 & 0.1 & 0.1 & 0.0 & 0.2 & 0.1 & 0.2 & 0.1 & 0.3 & \textbf{30.7} & 0.2 & 1.7 & 1.1 & \textbf{7.6} & 1.1 & 2.2 & 0.6 \\
    GA & 0.0 & \textbf{52.6} & 0.2 & 2.0 & 0.2 & 0.2 & 2.2 & 0.4 & 0.9 & 0.7 & 0.4 & 1.1 & 1.5 & 1.5 & 0.4 & 0.7 & 0.0 & 0.2 & 0.7 & 0.0 & 0.0 & 0.0 & 0.2 & \textbf{17.4} & 0.2 & 0.2 & 0.2 & \textbf{12.0} & 2.4 & 1.1 & 0.4 \\
    HI & 0.3 & \textbf{58.9} & 0.6 & 0.0 & 0.3 & 0.6 & 0.9 & 0.3 & 1.2 & 0.0 & 0.0 & 0.9 & 0.0 & 0.0 & 0.3 & 0.3 & 0.0 & 0.0 & 0.3 & 0.0 & 0.9 & 0.9 & 0.0 & \textbf{14.8} & 0.0 & 0.9 & 0.3 & \textbf{11.8} & 3.0 & 1.2 & 1.2 \\
    IA & 0.0 & \textbf{41.9} & 1.3 & 0.0 & 0.0 & 0.0 & 4.4 & 0.0 & 1.3 & 1.3 & 0.0 & 0.6 & 1.3 & 1.3 & 0.0 & 0.0 & 0.0 & 0.0 & 0.0 & 0.0 & 0.0 & 0.0 & 0.0 & \textbf{23.1} & 0.0 & 2.5 & 0.0 & \textbf{12.5} & \textbf{6.3} & 1.9 & 0.6 \\
    ID & 0.0 & \textbf{29.6} & 0.7 & 0.0 & 0.7 & 0.0 & 0.7 & 0.7 & 0.0 & 0.0 & 0.0 & 2.1 & 0.7 & 1.4 & 0.0 & 0.7 & 0.0 & 0.0 & 0.0 & 0.0 & 0.7 & 0.0 & 0.0 & \textbf{38.0} & 0.7 & 1.4 & 0.7 & \textbf{16.2} & 2.1 & 2.8 & 0.0 \\
    IL & 0.5 & \textbf{46.9} & 0.5 & 0.5 & 0.2 & 0.5 & 2.3 & 0.7 & 0.7 & 1.6 & 0.0 & 0.9 & 0.5 & 0.9 & 0.0 & 0.2 & 0.2 & 0.5 & 0.2 & 0.0 & 0.2 & 0.0 & 0.5 & \textbf{21.4} & 0.0 & 2.3 & 1.9 & \textbf{12.4} & 1.4 & 0.9 & 0.9 \\
    IN & 0.6 & \textbf{39.4} & 0.6 & 0.5 & 0.0 & 1.0 & 2.1 & 0.6 & 1.7 & 0.8 & 0.1 & 1.4 & 0.4 & 1.2 & 0.3 & 0.6 & 0.2 & 0.5 & 0.1 & 0.0 & 0.1 & 0.1 & 0.6 & \textbf{22.5} & 0.1 & 1.6 & 1.0 & \textbf{15.1} & 4.0 & 2.5 & 0.6 \\
    KS & 0.0 & \textbf{41.9} & 0.5 & 1.0 & 0.0 & 1.4 & 1.4 & 0.0 & 1.0 & 0.0 & 0.0 & 0.0 & 1.4 & 1.4 & 0.0 & 0.0 & 0.0 & 0.5 & 1.0 & 0.0 & 0.0 & 0.0 & 0.0 & \textbf{32.9} & 0.0 & 0.0 & 0.5 & \textbf{7.1} & \textbf{6.7} & 1.4 & 0.0 \\
    KY & 0.0 & \textbf{31.3} & 2.4 & 0.0 & 0.0 & 1.2 & 1.8 & 1.2 & 0.6 & 0.0 & 0.0 & 1.2 & 1.2 & 0.0 & 0.0 & 1.2 & 0.0 & 0.6 & \textbf{6.0} & 0.0 & 0.0 & 0.0 & 0.0 & \textbf{16.3} & 0.0 & 0.0 & 1.2 & \textbf{25.3} & 4.2 & 4.2 & 0.0 \\
    LA & 0.0 & \textbf{29.2} & 0.9 & 0.9 & 0.3 & 0.3 & 0.9 & 0.9 & 2.4 & 0.6 & 0.0 & 1.2 & 0.6 & 2.1 & 0.3 & 0.0 & 0.6 & 0.0 & 0.0 & 0.0 & 0.3 & 0.0 & 0.0 & \textbf{21.1} & 0.0 & 2.1 & 0.9 & \textbf{32.2} & 0.9 & 1.2 & 0.0 \\
    MA & 0.0 & \textbf{34.7} & 0.7 & 1.2 & 0.0 & 0.5 & 3.0 & 0.7 & 0.9 & 0.5 & 0.2 & \textbf{9.3} & 1.2 & 0.5 & 0.0 & 0.0 & 0.0 & 0.0 & 1.2 & 0.0 & 0.0 & 0.0 & 0.2 & \textbf{27.4} & 0.0 & 2.1 & 2.3 & \textbf{9.1} & 1.6 & 2.1 & 0.7 \\
    MD & 0.4 & \textbf{44.1} & 1.3 & 0.0 & 0.0 & 0.0 & 2.6 & 0.0 & 1.8 & 1.8 & 0.0 & 0.4 & 0.9 & 0.4 & 0.0 & 0.0 & 0.0 & 0.0 & 0.4 & 0.0 & 0.9 & 0.4 & 0.0 & \textbf{19.8} & 0.0 & 3.1 & 0.9 & \textbf{13.7} & \textbf{5.7} & 0.9 & 0.4 \\
    ME & 0.0 & \textbf{36.6} & 0.0 & 0.0 & 0.0 & 1.0 & 1.0 & 0.0 & 2.1 & 0.5 & 0.0 & 0.0 & 1.0 & 2.1 & 0.0 & 0.5 & 0.5 & 0.0 & 1.0 & 0.0 & 0.0 & 0.0 & 1.0 & \textbf{30.9} & 0.0 & 1.6 & 2.6 & \textbf{11.0} & 4.2 & 2.1 & 0.0 \\
    MI & 0.2 & \textbf{54.7} & 0.5 & 0.3 & 0.0 & 0.3 & 1.6 & 0.0 & 0.5 & 0.2 & 0.2 & 1.0 & 0.6 & 1.1 & 0.0 & 0.3 & 0.0 & 0.3 & 0.2 & 0.0 & 0.0 & 0.0 & 0.3 & \textbf{21.2} & 0.0 & 1.5 & 2.3 & \textbf{10.7} & 0.6 & 1.1 & 0.3 \\
    MN & 1.0 & \textbf{38.3} & 0.0 & 0.0 & 0.5 & 1.0 & 2.5 & 0.5 & 2.0 & 0.5 & 0.0 & 0.5 & 0.5 & 0.5 & 0.0 & 0.5 & 0.0 & 0.5 & 0.5 & 0.0 & 0.0 & 0.0 & 0.0 & \textbf{25.4} & 0.0 & 2.0 & 4.5 & \textbf{14.9} & 2.5 & 1.5 & 0.0 \\
    MO & 0.0 & \textbf{40.6} & 0.0 & 1.6 & 0.0 & 1.6 & 2.3 & 0.0 & 1.6 & 0.8 & 0.0 & 1.6 & 2.3 & 0.8 & 1.6 & 0.8 & 0.0 & 0.0 & 0.8 & 0.0 & 0.8 & 0.0 & 0.0 & \textbf{22.7} & 0.0 & 0.8 & 2.3 & \textbf{9.4} & 3.1 & 4.7 & 0.0 \\
    MS & 0.0 & \textbf{37.7} & 0.8 & 0.0 & 0.0 & 0.0 & 0.8 & 0.0 & 0.0 & 1.6 & 0.0 & 2.5 & 3.3 & 3.3 & 0.0 & 0.0 & 0.0 & 0.0 & 2.5 & 0.0 & 0.0 & 0.0 & 0.0 & \textbf{24.6} & 0.0 & 0.8 & 0.0 & \textbf{15.6} & 0.0 & 5.7 & 0.8 \\
    MT & 0.0 & \textbf{37.5} & 1.3 & 1.3 & 0.0 & 0.0 & \textbf{6.3} & 0.0 & 2.5 & 3.8 & 0.0 & 0.0 & 0.0 & 0.0 & 0.0 & 0.0 & 0.0 & 0.0 & 3.8 & 0.0 & 0.0 & 0.0 & 1.3 & \textbf{25.0} & 0.0 & 1.3 & 0.0 & \textbf{13.8} & 1.3 & 1.3 & 0.0 \\
    NC & 0.2 & \textbf{36.5} & 0.9 & 0.6 & 0.2 & 1.1 & 2.1 & 0.2 & \textbf{9.1} & 0.2 & 0.0 & 0.6 & 1.5 & 1.1 & 0.0 & 0.2 & 0.0 & 0.0 & 0.0 & 0.0 & 0.2 & 0.0 & 0.4 & \textbf{28.7} & 0.0 & 0.4 & 0.2 & \textbf{14.2} & 0.9 & 0.4 & 0.2 \\
    ND & 0.0 & \textbf{36.3} & 0.4 & 0.0 & 0.0 & 0.8 & 3.4 & 1.1 & 0.8 & 0.8 & 0.4 & 1.5 & 0.4 & 0.8 & 0.0 & 0.8 & 0.0 & 0.0 & 0.0 & 0.0 & 0.4 & 0.0 & 0.4 & \textbf{23.3} & 0.0 & 1.1 & 1.5 & \textbf{21.4} & 1.9 & 2.3 & 0.4 \\
    NE & 0.0 & \textbf{35.7} & 0.9 & 1.8 & 0.0 & 1.3 & 3.1 & 0.9 & 1.8 & 0.4 & 0.0 & 0.0 & 0.9 & 0.9 & 0.4 & 0.9 & 0.0 & 0.0 & 0.9 & 0.0 & 0.4 & 0.0 & 0.4 & \textbf{27.2} & 0.0 & 1.8 & 0.9 & \textbf{11.6} & 2.2 & 4.9 & 0.4 \\
    NH & 0.8 & \textbf{31.5} & 0.8 & 1.6 & 0.0 & 0.0 & 0.8 & 0.0 & 1.6 & 0.8 & 0.0 & \textbf{24.4} & 0.8 & 0.8 & 0.0 & 0.8 & 0.8 & 0.0 & 0.0 & 0.0 & 0.0 & 0.0 & 0.0 & \textbf{20.5} & 0.0 & 1.6 & 0.8 & \textbf{7.9} & 2.4 & 0.8 & 0.8 \\
    NJ & 1.6 & \textbf{54.9} & 1.2 & 0.3 & 0.2 & 0.5 & 1.7 & 0.9 & 2.4 & 0.7 & 0.0 & 0.3 & 0.7 & 0.9 & 0.0 & 0.5 & 0.0 & 0.0 & 0.9 & 0.0 & 0.2 & 0.2 & 0.7 & \textbf{17.8} & 0.0 & 1.2 & 0.3 & \textbf{8.3} & 1.7 & 1.6 & 0.3 \\
    NM & 0.0 & \textbf{19.3} & \textbf{6.8} & 0.0 & 1.1 & 0.0 & 3.4 & 0.0 & 1.1 & 0.0 & 0.0 & 1.1 & 0.0 & 2.3 & 0.0 & 1.1 & 0.0 & 0.0 & 0.0 & 0.0 & 0.0 & 0.0 & 0.0 & \textbf{21.6} & 1.1 & \textbf{15.9} & \textbf{6.8} & \textbf{13.6} & 3.4 & 1.1 & 0.0 \\
    NV & 1.1 & \textbf{46.1} & 0.6 & 0.6 & 1.1 & 0.6 & 2.2 & 1.1 & 1.7 & 1.1 & 0.0 & 0.6 & 1.7 & 2.2 & 0.0 & 0.6 & 0.0 & 0.0 & 1.1 & 0.0 & 0.6 & 0.6 & 0.0 & \textbf{16.1} & 0.0 & 0.6 & 0.6 & \textbf{13.3} & 3.3 & 0.6 & 2.2 \\
    NY & 0.7 & \textbf{49.7} & 0.4 & 0.5 & 0.2 & 0.5 & 0.6 & 0.3 & 0.7 & 0.3 & 0.1 & 0.7 & 0.5 & 1.2 & 0.1 & 0.3 & 0.1 & 0.1 & 0.2 & 0.0 & 0.1 & 0.2 & 0.2 & \textbf{21.8} & 0.1 & 1.0 & 0.6 & \textbf{16.2} & 1.7 & 0.9 & 0.2 \\
    OH & 0.2 & \textbf{37.3} & 0.7 & 0.7 & 0.0 & 0.9 & 2.2 & 0.7 & 1.3 & 0.4 & 0.2 & 1.7 & 0.4 & 1.5 & 0.0 & 0.2 & 0.0 & 0.0 & 0.9 & 0.2 & 0.2 & 0.7 & 0.7 & \textbf{24.8} & 0.0 & 4.1 & 1.5 & \textbf{7.6} & \textbf{6.8} & 3.1 & 1.1 \\
    OK & 0.0 & \textbf{35.4} & 0.6 & 0.0 & 0.0 & 0.6 & 1.2 & 0.0 & 0.6 & 0.6 & 0.0 & 1.9 & 1.9 & 0.6 & 0.0 & 0.6 & 0.0 & 0.0 & 0.6 & 0.0 & 0.6 & 0.0 & 0.0 & \textbf{29.8} & 0.0 & 1.2 & 0.6 & \textbf{15.5} & 4.3 & 1.9 & 1.2 \\
    OR & 0.3 & \textbf{32.5} & 2.4 & 0.8 & 0.3 & 1.8 & 2.1 & 0.3 & 0.8 & 1.3 & 0.0 & 0.8 & 1.0 & 1.0 & 0.3 & 0.0 & 0.0 & 0.0 & 0.3 & 0.0 & 0.0 & 0.0 & 0.5 & \textbf{32.5} & 0.3 & 2.4 & 1.3 & \textbf{12.6} & 3.1 & 1.6 & 0.0 \\
    PA & 0.5 & \textbf{29.3} & 1.0 & 0.3 & 0.2 & 0.7 & 2.0 & 0.7 & 0.7 & 0.2 & 0.2 & 1.0 & 0.5 & 0.3 & 0.0 & 0.0 & 0.0 & 0.3 & 0.5 & 0.0 & 0.0 & 0.2 & 0.3 & \textbf{20.0} & 0.0 & 1.5 & 3.0 & \textbf{16.9} & \textbf{17.9} & 1.2 & 1.0 \\
    RI & 0.4 & \textbf{60.0} & 0.0 & 0.0 & 0.0 & 0.7 & 1.1 & 0.4 & 0.7 & 0.4 & 0.0 & 1.1 & 1.4 & 0.0 & 0.0 & 0.4 & 0.0 & 0.0 & 0.0 & 0.0 & 0.0 & 0.0 & 0.4 & \textbf{20.4} & 0.4 & 0.0 & 4.2 & 3.9 & 2.1 & 2.1 & 0.4 \\
    SC & 0.5 & \textbf{49.5} & 1.1 & 0.0 & 0.0 & 1.6 & 1.6 & 0.5 & 2.1 & 0.0 & 0.0 & 1.6 & 1.1 & 0.0 & 0.0 & 0.5 & 0.0 & 0.0 & 0.0 & 0.0 & 0.0 & 0.0 & 0.0 & \textbf{23.9} & 0.5 & 0.5 & 0.5 & \textbf{10.6} & 2.1 & 1.6 & 0.0 \\
    SD & 0.0 & \textbf{31.3} & 2.1 & 0.0 & 0.0 & 2.1 & 0.0 & 0.0 & 0.0 & 2.1 & 0.0 & 0.0 & 0.0 & 2.1 & 0.0 & 0.0 & 0.0 & 0.0 & 2.1 & 0.0 & 0.0 & 0.0 & 0.0 & \textbf{43.8} & 0.0 & 2.1 & 2.1 & \textbf{6.3} & 4.2 & 0.0 & 0.0 \\
    TN & 0.0 & \textbf{52.4} & 1.0 & 0.0 & 0.0 & 1.0 & 1.0 & 0.5 & 0.5 & 0.5 & 0.0 & 0.5 & 1.0 & 1.4 & 0.0 & 1.0 & 0.0 & 0.0 & 0.0 & 0.0 & 1.0 & 0.0 & 0.0 & \textbf{19.2} & 0.0 & 2.4 & 1.9 & \textbf{10.6} & 2.4 & 1.4 & 0.5 \\
    TX & 0.4 & \textbf{42.0} & 0.8 & 0.0 & 0.1 & 1.0 & 3.0 & 0.3 & 0.4 & 1.5 & 0.4 & \textbf{8.1} & 1.1 & 1.9 & 0.1 & 0.5 & 0.2 & 0.0 & 0.4 & 0.0 & 0.3 & 0.0 & 0.4 & \textbf{22.4} & 0.1 & 1.9 & 1.1 & \textbf{8.9} & 1.3 & 1.3 & 0.4 \\
    UT & 1.1 & \textbf{39.5} & 2.6 & 0.0 & 0.0 & 0.0 & 2.1 & 1.1 & 1.1 & 2.1 & 0.0 & 0.5 & 2.1 & 1.6 & 0.5 & 0.5 & 1.1 & 0.5 & 1.6 & 0.0 & 0.5 & 0.5 & 1.1 & \textbf{23.2} & 0.5 & 2.6 & 1.1 & \textbf{7.4} & 1.1 & 2.1 & 2.1 \\
    VA & 0.5 & \textbf{49.0} & 1.6 & 0.5 & 0.3 & 0.3 & 2.7 & 0.3 & 0.5 & 0.5 & 0.0 & 0.5 & 1.1 & 1.9 & 0.0 & 1.1 & 0.0 & 0.3 & 0.5 & 0.0 & 0.0 & 0.0 & 0.0 & \textbf{22.5} & 0.3 & 0.5 & 1.1 & \textbf{9.3} & 1.9 & 2.5 & 0.0 \\
    VT & 0.0 & \textbf{28.6} & 0.0 & 0.0 & 0.0 & 0.0 & 1.3 & 0.0 & 0.0 & 0.0 & 0.0 & 0.0 & 1.3 & 0.0 & 0.0 & 0.0 & 0.0 & 0.0 & 1.3 & 0.0 & 1.3 & 0.0 & 0.0 & \textbf{32.5} & 0.0 & 1.3 & 0.0 & \textbf{15.6} & \textbf{11.7} & \textbf{5.2} & 0.0 \\
    WA & 0.3 & \textbf{42.1} & 1.1 & 0.9 & 0.2 & 0.5 & 0.9 & 0.5 & 2.6 & 0.9 & 0.0 & 0.6 & 0.6 & 0.9 & 0.3 & 0.8 & 0.2 & 0.3 & 0.5 & 0.0 & 0.2 & 0.0 & 0.8 & \textbf{24.8} & 0.2 & 1.8 & 0.8 & \textbf{13.4} & 1.8 & 2.3 & 0.2 \\
    WI & 1.0 & \textbf{41.1} & 1.0 & 2.4 & 0.3 & 1.0 & 2.1 & 0.3 & 1.0 & 0.0 & 0.0 & 2.4 & 1.4 & 2.1 & 0.7 & 0.7 & 0.3 & 0.0 & 0.0 & 0.0 & 0.0 & 0.0 & 0.3 & \textbf{22.6} & 1.4 & 2.1 & 1.4 & \textbf{8.6} & 3.4 & 2.1 & 0.3 \\
    WV & 0.0 & \textbf{29.9} & 0.0 & 1.5 & 0.0 & 1.5 & 3.0 & 0.0 & 0.0 & 1.5 & 0.0 & 0.0 & 1.5 & 0.0 & 1.5 & 1.5 & 0.0 & 1.5 & 3.0 & 0.0 & 0.0 & 0.0 & 0.0 & \textbf{25.4} & 0.0 & 4.5 & 1.5 & \textbf{16.4} & 4.5 & 1.5 & 0.0 \\
    WY & 0.0 & \textbf{32.6} & 2.2 & 0.0 & 0.0 & 0.0 & 4.3 & 0.0 & 0.0 & 0.0 & 0.0 & 0.0 & 0.0 & 0.0 & 0.0 & 0.0 & 2.2 & 2.2 & 0.0 & 0.0 & 4.3 & 4.3 & 0.0 & \textbf{34.8} & 0.0 & 0.0 & 2.2 & \textbf{6.5} & 4.3 & 0.0 & 0.0 \\
    \bottomrule
    \vspace*{1\baselineskip}
    \end{tabular}%
  \caption*{\textit{Note:} (in original order)\\
  \begin{CJK}{UTF8}{gkai} 皖: Anhui,
  京:Beijing,  渝: Chongqing,  闽: Fujian,  甘: Gansu, 粤: Guangdong, 桂: Guangxi, 贵: Guizhou, 琼: Hainan,\\ 冀: Hebei, 黑: Heilongjiang, 豫: Henan, 鄂: Hubei, 湘: Hunan, 蒙: Inner Mongolia, 苏: Jiangsu, 赣: Jiangxi, 吉: Jilin,\\ 辽: Liaoning, 宁: Ningxia, 青: Qinghai, 陕: Shaanxi, 鲁: Shandong, 沪: Shanghai, 晋: Shanxi, 川: Sichuan, 津: Tianjin,\\ 藏: Tibet, 新: Xinjiang, 云: Yunnan, 浙: Zhejiang.\end{CJK} }
  \label{tab:addlabel}%
\end{table*}%

Two immediate observations follow. First, nationwide most of the tweets can be attributed to three Chinese provinces: Beijing, Shanghai and Tibet. Xinjiang province makes a distant fourth. Second, there exists large inter-state variation. For example, the share of tweets that go to Beijing ranges from 24.3\% for Delaware to 60.0\% for Rhode Island. For the majority of the U.S. states (44 out of 51), the largest share of tweets goes to Beijing.\footnote{We observe two ties and in both cases we decide to side with Beijing.} For Delaware, Idaho, New Mexico, South Dakota, Vermont and Wyoming (6 out of 51), Shanghai receives the most tweets. For Louisiana, it is Tibet. 


\subsection{Friendliness and Variance}
Opinion polls on U.S.-China relations mostly stop at the national level. By contrast, we are able to explore state level nuances. In particular, we are able to measure the friendliness of each state towards China. This is important because hardly ever have international relations been measured without regard to security issues, and security concerns vary across countries. Studying state level attitudes enables us to control for security issues in a perfect way as we can assume all U.S. states share the same security concerns with regards to China. In addition to friendliness, we are also able to measure how varied perceptions of China are in each state.

\begin{figure*}[!htbp]
\includegraphics[height=10cm, width=18cm]{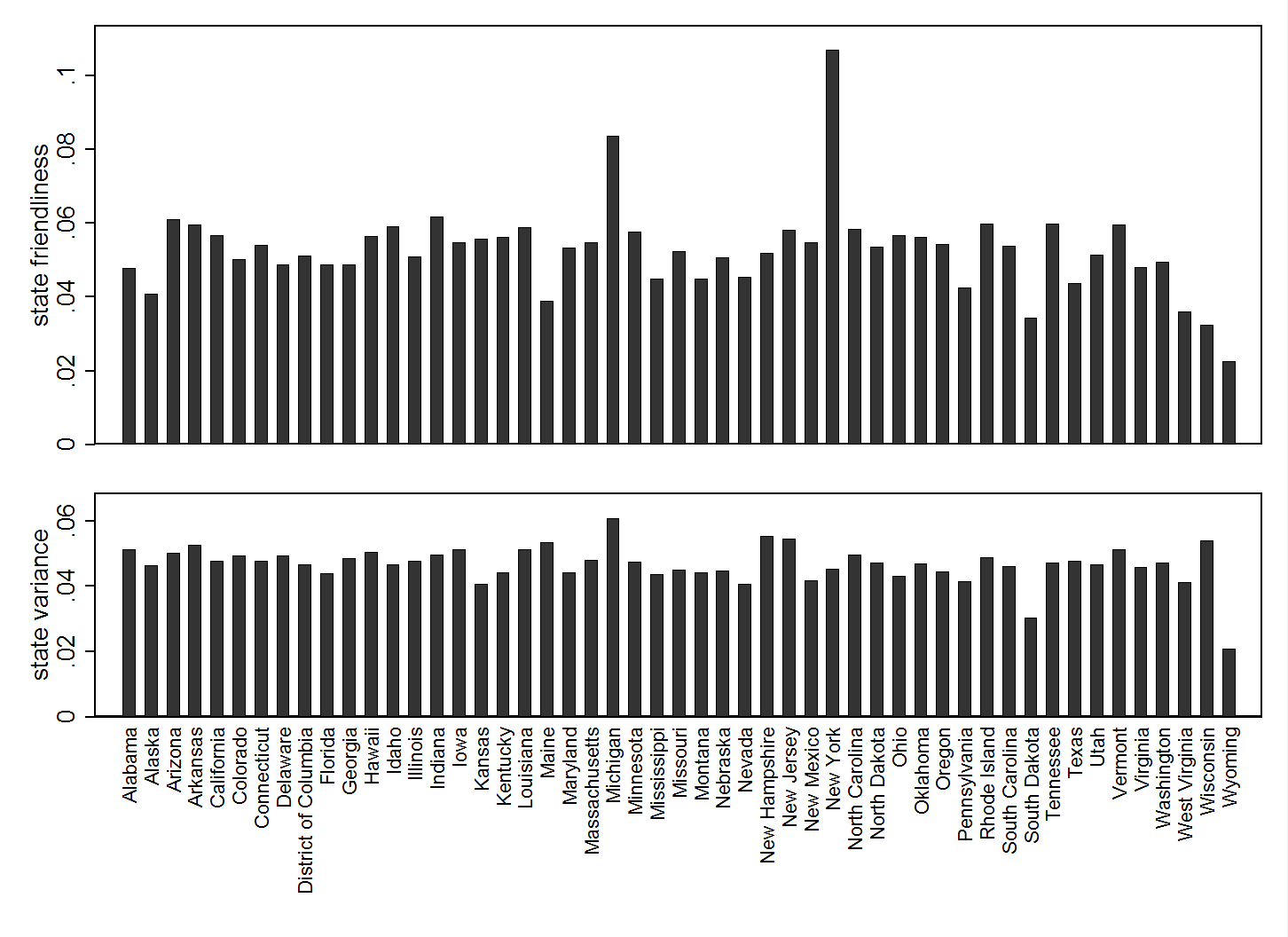}
\caption{Friendliness and Variance. The top figure reports the friendliness index for each state. The bottom figure reports the variance index for each state.}
\end{figure*}

The results are reported in Fig. 5. We find that the top four most friendly states are New York, Michigan, Indiana and Arizona and that the least friendly four states are Wyoming, Wisconsin, South Dakota and West Virginia. In terms of variance, we find the top four most homogeneous states are Wyoming, South Dakota, Kansas and Nevada and that the least homogeneous states are Michigan, New Hampshire, New Jersey and Wisconsin. 

Though we do not explore it here, we point out that it will be of great significance to  analyze the causal mechanisms behind these differences between states. We suggest that international trade and Chinese immigration can be influential factors.

\section{Individual level Analysis}
In this section, we first introduce our statistical model and then report our main estimation results. Lastly, we perform two separate \textit{F} tests on the state and date control variables.

\subsection{The Model}
Our primary goal here is to replicate some of the stylized results from opinion polls \cite{OpinionLeaders,StudyChina}:

\begin{itemize}
  \item American individuals who follow news about China have less favorable views of China.
  \item American opinion leaders are more likely to have a favorable view of China.
  \item American students who have studied in China are more likely to have a favorable view of China.
\end{itemize}

Our dependent variable is the polarity of the tweet. We use \textbf{followers} and \textbf{intensity}, introduced in Section 2, to capture the effects of opinion leaders. That is, we treat individuals with more followers and more posted tweets as opinion leaders. In the original study, the opinion leaders consist of  U.S. government officials, think tank leaders, media personnel, business executives, and university faculties \cite{OpinionLeaders}. We use variables \textbf{followees} and \textbf{retweet} to capture the effects of following news. The variable \textbf{reply} controls for the effects of being in a conversation.

Additionally, we examine the effects of Chinese ethnicity on attitudes towards China. The size of our data set also allows us to control for state and time fixed effects. As previous sections have shown, there is large variation across U.S. states and between different days, so we decide to control for both state and time effects. Altogether, this suggests that following statistical model:
\begin{align*}
     y_i&=\beta_1\cdot followers_i+\beta_2\cdot followees_i+\beta_3\cdot  retweet_i\\
        &+\beta_4\cdot  reply_i+\beta_5\cdot  experience_i+\beta_6\cdot  intensity_i\\
        &+D_i\boldsymbol\alpha+C_i\boldsymbol\lambda+S_i\boldsymbol\gamma+\epsilon_i
\end{align*}

where ${D_i}=[{d_1^i\:d_2^i\: ...\:d_{17}^i}]$ control for date effects, ${S_i}=[{s_1^i\:s_2^i\: ...\:s_{50}^i}]$ control for state effects and  ${C_i}$ controls for the   effects of Chinese ethnicity.

\subsection{The Main Results}
We estimate our model with OLS regression and the estimation results are reported in Table 4. Each column represents one regression and has its own specification. The first column displays estimates without controlling for state and time fixed effects. The second column presents the same coefficients but controls for state fixed effects. The third column controls for time fixed effects. The fourth column incorporates both state and time fixed effects. 

For the fifth column, we aggregate the tweets on an individual and daily basis by taking the average. This reduces the number of observations from 245,664 to 162,982. The six column, with 242,673 observations, reports estimates for individuals not identified as ethnically Chinese. The seventh column, with 2991 observations, reports estimates for individuals identified as ethnically Chinese. 

\begin{table*}[!htbp]\centering
\caption{Estimation Results}
\def\sym#1{\ifmmode^{#1}\else\(^{#1}\)\fi}
  \tabcolsep=0.085cm

\begin{tabular}{l*{7}{c}}
\hline\hline
			&\multicolumn{7}{c}{tweet polarity}\\\cline{2-8} 
            &\multicolumn{1}{c}{(1)}&\multicolumn{1}{c}{(2)}&\multicolumn{1}{c}{(3)}&\multicolumn{1}{c}{(4)}&\multicolumn{1}{c}{(5)}&\multicolumn{1}{c}{(6)}&\multicolumn{1}{c}{(7)}\\
            &\multicolumn{1}{c}{Baseline}&\multicolumn{1}{c}{State effects}&\multicolumn{1}{c}{Date effects}&\multicolumn{1}{c}{State \& Date effects}&\multicolumn{1}{c}{Aggregated}&\multicolumn{1}{c}{Non-Chinese Ethnicity}&\multicolumn{1}{c}{Chinese Ethnicity}\\
\hline
followers   &    2.80e-11         &   -1.33e-09         &    1.62e-10         &   -1.21e-09         &   -9.00e-11         &   -1.23e-09         & 0.000000725\sym{**} \\
            &      (0.02)         &     (-0.79)         &      (0.10)         &     (-0.72)         &     (-0.04)         &     (-0.73)         &      (2.73)         \\
followees   &-0.000000353\sym{***}&-0.000000333\sym{***}&-0.000000352\sym{***}&-0.000000331\sym{***}&-0.000000154\sym{**} &-0.000000326\sym{***}& -0.00000138\sym{**} \\
            &     (-8.22)         &     (-7.74)         &     (-8.19)         &     (-7.70)         &     (-2.80)         &     (-7.52)         &     (-3.04)         \\
retweet     &-0.000000907\sym{**} &-0.000000871\sym{**} &-0.000000935\sym{**} &-0.000000897\sym{**} & -0.00000154\sym{***}&-0.000000898\sym{**} &  0.00000340         \\
            &     (-2.94)         &     (-2.83)         &     (-3.03)         &     (-2.91)         &     (-4.75)         &     (-2.91)         &      (0.18)         \\
reply       &      0.0219\sym{***}&      0.0217\sym{***}&      0.0219\sym{***}&      0.0217\sym{***}&      0.0219\sym{***}&      0.0215\sym{***}&      0.0192         \\
            &     (17.10)         &     (16.90)         &     (17.11)         &     (16.91)         &     (14.41)         &     (16.64)         &      (1.73)         \\
experience         &  0.00000337\sym{***}&  0.00000293\sym{***}&  0.00000332\sym{***}&  0.00000287\sym{***}&  0.00000170\sym{*}  &  0.00000290\sym{***}&-0.000000992         \\
            &      (5.68)         &      (4.91)         &      (5.60)         &      (4.81)         &      (2.46)         &      (4.82)         &     (-0.17)         \\
intensity   &   0.0000930\sym{***}&   0.0000870\sym{***}&   0.0000927\sym{***}&   0.0000865\sym{***}&   0.0000855\sym{***}&   0.0000869\sym{***}&   -0.000113         \\
            &     (78.96)         &     (67.11)         &     (78.46)         &     (66.54)         &     (40.29)         &     (66.62)         &     (-1.85)         \\
Chinese     &    0.000187         &     0.00337         &    0.000348         &     0.00344         &      0.0210\sym{***}&                     &                     \\
            &      (0.05)         &      (0.83)         &      (0.09)         &      (0.85)         &      (3.98)         &                     &                     \\
state effects  &  No&   Yes& No   &   Yes  &   Yes &   Yes  &   Yes\\
date effects  &  No&   No&  Yes   &   Yes  &   Yes &   Yes  &   Yes\\
constant  &   Yes&   Yes&Yes   &   Yes  &   Yes &   Yes  &   Yes\\
\hline
\(N\)       &      245664         &      245664         &      245664         &      245664         &      162982         &      242673         &        2991         \\
adj. \(R^{2}\)&       0.028         &       0.030         &       0.030         &       0.031         &       0.016         &       0.031         &       0.042         \\
\hline\hline

\multicolumn{8}{l}{\footnotesize \textit{t} statistics in parentheses}\\
\multicolumn{8}{l}{\footnotesize \sym{*} \(p<0.05\), \sym{**} \(p<0.01\), \sym{***} \(p<0.001\)}\\
\multicolumn{8}{l}{\footnotesize Estimates for control variables are not reported. In the following subsection, joint \textit{F} tests are used to show that they are statistically different from zero.}\\
\end{tabular}
\end{table*}

Our results (Columns 1-5) are consistent with the findings cited above. Specifically, the coefficient for \textbf{intensity} is positive and statistically significant, indicating that opinion leaders, in our case individuals who tweet more often, are more likely to have a positive view of China. The coefficient for \textbf{followers} is not statistically significant. 

Coefficients for \textbf{followees} and \textbf{retweet} are both negative and statistically significant. This is consistent with the finding that individuals who follow news about China have less favorable views of the country.

 We also find \textbf{experience} to have a statistically positive effect. This suggests a positive learning experience using Twitter and supports that finding that American students who have studied in China tend to have more positive views of the country. 
  
  The estimate for \textbf{Chinese} is positive in all the first five specifications, but is statistically significant only in Test 5, when we aggregate individuals' daily tweets. 

Comparing the results in the sixth column and the seventh column, we find individuals of Chinese ethnicity behave differently from the rest of the sample. For example, the coefficient on \textbf{followers} is positive and statistically significant. The estimates on \textbf{retweet}, \textbf{reply}, \textbf{experience} and \textbf{intensity} are no longer statistically significant.
\subsection{\textit{F} tests} 
  To test the significance of the state and date effects, we perform two separate \textit{F} tests based on the fourth column in Table 4. The testing results are reported below. With F(50, 245589)=6.88, we reject the null hypothesis that coefficients on state control variables are all zeros. With F(17, 245589)=16.99, we are able to reject the null hypothesis that the coefficients on date control variables are all zeros. These tests further confirm the existence of inter-state and inter-day variations.
  
\begin{table}[!h]
\centering
\caption{\textit{F} Tests on control variables}
\label{my-label}
\begin{tabular}{lll}
\hline
state coefficients & F(50, 245589)=6.88  & Prob\textgreater F=0.0000 \\
date coefficients  & F(17, 245589)=16.99 & Prob\textgreater F=0.0000                \\
\hline
\end{tabular}
\end{table}

\section{Conclusion}
The U.S.-China relationship is arguably the most important bilateral relationship in the 21st century. It demands detailed measurement and analysis. In this paper, we have proposed a new method to measure this relationship using tweets. With a large data set, we are able to carry out state level analysis. Utilizing geo-information and time stamps, we can control for fixed state and time effects. We demonstrate the existence of inter-state and inter-day variations and control for them in our regression analysis.

Our work replicates some stylized results from opinion polls as well as generate new insights. At the state level, we find New York, Michigan, Indiana and Arizona are the top four most China-friendly states. Wyoming, South Dakota, Kansas and Nevada are most homogeneous. At the individual level, we find attitudes towards China improve as the individuals' Twitter experience grows longer and more intense. We also find individuals of Chinese ethnicity are more China-friendly.

Our study is the first to analyze U.S.-China relations using Twitter data. The results we achieve are very encouraging. In future research we intend to increase the size of our data set and further refine our analytic tools.

\section*{Acknowledgment}
Yu Wang would like to thank the Department of Political Science at the University of Rochester for warm encouragement and generous funding. This work was also generously supported in part by Google, Yahoo,  Adobe, TCL, and New York State CoE CEIS and IDS.



%


\bibliographystyle{IEEEtran}
%
\bibliography{hello}
\end{document}